\documentclass[12pt]{article}
\usepackage{graphicx}
\usepackage{amsmath}
\usepackage{amsfonts}
\usepackage{authblk}
\usepackage{mathtools}
\usepackage{footnote}
\usepackage{bm}
\usepackage{scalerel,stackengine,longtable}
\stackMath
\newcommand\reallywidehat[1]{%
\savestack{\tmpbox}{\stretchto{%
  \scaleto{%
    \scalerel*[\widthof{\ensuremath{#1}}]{\kern-.6pt\bigwedge\kern-.6pt}%
    {\rule[-\textheight/2]{1ex}{\textheight}}%WIDTH-LIMITED BIG WEDGE
  }{\textheight}% 
}{0.5ex}}%
\stackon[1pt]{#1}{\tmpbox}%
}
\newcommand{\Lim}[1]{\raisebox{0.5ex}{\scalebox{0.8}{$\displaystyle \lim_{#1}\;$}}}

\newcommand\eref[1]{(\ref{#1})}
\renewcommand{\theequation}{\arabic{equation}}
\newtheorem{theorem}{Theorem}

\newtheorem{lemma}{Lemma}

\def\bmc{\mbox{\boldmath $c$}}

\def\bmR{\mbox{\boldmath $R$}}
\def\bmS{\mbox{\boldmath $S$}}
\def\bmT{\mbox{\boldmath $T$}}

\def\bmY{\mbox{\boldmath $Y$}}

\def\bmx{\mbox{\boldmath $x$}}

\def\bm1{\mbox{\boldmath $1$}}

\def\bmbeta{\mbox{\boldmath $\beta$}}

\def\bmvep{\mbox{\boldmath $\varepsilon$}}

\def\Ripij{\mbox{$R_{ij, ii'}$}}
\def\Ripi1{\mbox{$R_{i1, ii'}$}}
\def\Ripini{\mbox{$R_{in_i, ii'}$}}

\def\Riip1{\mbox{$R_{i'1, ii'}$}}
\def\Riipnip{\mbox{$R_{i'n_{i'}, ii'}$}}
\def\bmRipi{\mbox{{\boldmath$R$}$_{ii'}$}}
\def\Xii{\mbox{ $X_{ii'}$}}
\def\X1ii{\mbox{ $X_{1,ii'}$}}

\def\x0ii{\mbox{$\boldmath{x}_{0,ii'}$}} 

\def\H1ii{\mbox{ $H_{1,ii'}$}}
\def\x1ij{\mbox{$x_{1,ij}$}}
\def\xpij{\mbox{$x_{p,ij}$}}

\begin{document}
\begin{titlepage}
	
\title{\Large Simultaneous Rank Tests in Analysis of Covariance Based on Pairwise Ranking}
\date{}
%\date{12/8/2017}

\author{
 Hossein Mansouri\thanks{Corresponding author, email: hossein.mansouri@ttu.edu}\  \  and Fangyuan Zhang}
\affil{\footnotesize Department of Mathematics and Statistics, Texas Tech University, Lubbock, TX 79409-1042, USA\\
}
\maketitle

\end{titlepage}

%%%%%%%%%%%%%%%%%%%%%%%%%%%%%%%%%%%%%%%%%%%%%%%%%%%%%%%%%%%%%%%%%%%%%%%

\begin{abstract}
	Nonparametric tests provide robust and powerful alternatives to the corresponding least squares methods. There are two approaches to nonparametric pairwise comparisons of treatment effects, the method based on pairwise rankings and the method based on overall ranking. The former is generally recommended in the literature because of its strong control of familywise error rate. However, this method is developed only for one-way layouts and randomized complete blocks. By combining the method of aligned ranks and pairwise ranking, we extend the Steel-Dwass pairwise comparisons to the analysis of covariance  and factorial models for both one-sided and two-sided comparisons as well as testing for treatment versus control. Unlike the traditional two-sample standardization of test statistics, we propose a weighted estimate of the scale parameter for ranks and show through simulation that it has superior small sample performance by controlling the familywise error rate at nominal level. This method provides an improvement for large sample approximation of Steel-Dwass method for one-way layouts. The marginal and joint asymptotic distributions are derived and power comparisons are made with the method of aligned rank transformation and the least squares method.   
\end{abstract}

\baselineskip=24pt \noindent \textbf{Key Words}: Aligned rank
transformation;  Multiple comparisons; Familywise error rate.


%%%%%%%%%%%%%%%%%%%%%%%%%%%%%%%%%%%%%%%%%%%%%%%%%%%%%%%%%%%%%%%%%
%%%%%%%%%%%%%%%%%%%%%%%%%%%%%%%%%%%%%%%%%%%%%%%%%%%%%%%%%%%%%%%%%

\section{Introduction}

Nonparametric methods of statistical analysis provide robust methods of testing and estimation. Simultaneous  pairwise comparisons based on ranks have been around for nearly as long as their parametric counterparts. For one-way layouts and randomized complete blocks, there are two nonparametric methods of pairwise comparisons of treatment effects, the method based on pairwise or separate rankings and the method based on overall ranking.

For balanced one-way layouts, nonparametric method of simultaneous pairwise comparisons based on the Wilcoxon \cite{wilcoxon}  rank-sum statistics obtained by  separate rankings of pairs of samples is due to Steel \cite{steel} who derived the exact distribution of the test statistics and generated exact tables of critical values for 3 treatments and $n= 2, 3, 4$ replications per treatment.  Dwass \cite{dwass} independently proposed the same test and studied the asymptotic distribution of the test statistic based on maximum modulus of standardized Wilcoxon rank-sum statistics. Extensions to unbalanced one-way layouts are due to Hochberg and Tamhane \cite{hochberg} and Critchlow and Fligner \cite{critchlow} who in addition proposed simultaneous confidence intervals based on the  two-sample  Hodges-Lehmann rank estimates of the location parameters, Hodges and Lehmann \cite{hodgesLehmann}. Simultaneous pairwise comparisons based on the overall ranking of the samples are due to Nemenyi \cite{nemenyi} and Dunn \cite{dunn}.

Each of the two methods mentioned above has its advantages and disadvantages, the general consensus in the literature is to recommend the method of separate rankings over the method of overall ranking since the former controls the familywise error rate (FWER) in strong sense and the latter controls it in weak sense, Hochberg and Tamhane \cite{hochberg} and Hsu \cite{hsu}. Briefly, weak control of FWER  means that it is controlled under the global null hypothesis that all group effects are equal. Strong control of FWER means that it is controlled under individual null hypotheses and different configurations of the hypotheses that imply the individual null hypothesis.

Although pairwise comparisons based on overall ranking are extended to factorial designs and general linear models based on the aligned rank transformation methodology, Mansouri \cite{mansouri15}, the methods based on separate rankings are limited to one-way layouts and randomized complete blocks, see \cite{hochberg}. It is the aims of this investigation to extend the method of pairwise comparisons  based on separate rankings to a model that includes the factorial models and analysis of covariance as a special case. This method combines the methods of aligned rank transformation and separate rankings. We show that the resulting pairwise comparisons procedure controls FWER in strong sense. Unlike one-way layouts and randomized complete blocks for which the resulting rank tests are distribution-free, rank tests for higher order designs are asymptotically distribution-free and hence control of FWER is in asymptotic sense. For small to moderate designs, we will propose approximation methods that control FWER at the nominal level.

In section 2 we formulate the problem and offer tests for simultaneous pairwise comparisons among treatments and simultaneous pairwise comparisons of treatments with a control. We present these tests for two-sided as well as one-sided alternatives. These tests are formulated as functions of general rank-scores based on pairwise rankings of aligned samples. We propose different estimates of the rank-score scale parameter and recommend an estimate that is a weighted average of two-sample estimates of the rank-score scale parameter that according to the simulations study of section 4  produces the most satisfactory results for controlling the FWER in small to moderate size data sets.  We show that the testing procedures reduce to the existing tests for one-way layouts when the Wilcoxon rank-scores are used. In section 3, we derive the asymptotic joint distribution of the vector of statistics for pairwise comparisons and the vector of statistics for treatments versus control comparisons. Often in the literature, the asymptotic results are used to develop testing procedures by using the asymptotic distribution of the test statistics. In section 4, in addition to the simulation study concerning robustness of validity in controlling the Type I FWER, we conduct power comparisons between the competing methods. Simulation results show that using weighted estimate of the scale parameter in pairwise ranking not only enjoys the robustness of validity  property, but it is often more powerful than the competing tests. In section 5, we apply the method to a real data set obtained from an epidemiological study available in the literature. Theoretical proofs are presented in the appendix.

%%%%%%%%%%%%%%%%%%%%%%%%%%%%%%%%%%%%%%%%%%%%%%%%%%%%%%%%%%%%%%%%%
%%%%%%%%%%%%%%%%%%%%%%%%%%%%%%%%%%%%%%%%%%%%%%%%%%%%%%%%%%%%%%%%%

\section{Problem Formulation and Pairwise Tests}

Let \mbox{$Y_{ij}$} denote random observations that follow the linear model

\begin{align}
\mbox{Y}_{ij}&=\mu_i+\beta_1\x1ij+\ldots+\beta_p\xpij+\varepsilon_{ij}&\nonumber\\
&=\mu_i+\bmx^\textsc{t} _{ij}\bmbeta+\varepsilon_{ij},& i=1, \ldots, g; j=1, \ldots, n_i \label{model}
 \end{align}

\noindent where $\mu_i$ denotes the effect of the $i$-th group,
$\bmx^\textsc{t} _{ij}=(\x1ij, \ldots, \xpij)$ is the vector of covariates associated with the $j$-th subject in the $i$-th group, $\bmbeta=(\beta_1,\ldots,\beta_p)^\textsc{t}$ is the vector of parameters associated with the covariates and $\varepsilon_{ij}$'s are independent random variables with a continuous cumulative distribution function (cdf) $F(\cdot)$ such that $F(0)=0.5$.

The covariates in \eref{model} can be quantitative or qualitative. For instance, the  model may correspond to factorial data where $\mu_i$'s  represent specific effects of interest and the remaining parameters are collectively considered as covariate effects.

The main objective of the present article is to conduct pairwise comparisons of the form
\begin{align}
&H_{0,ii'}:\ \ \mu_i-\mu_{i'}=0   \nonumber\\
&H_{1,ii'}:\ \ \mu_i-\mu_{i'}\ne 0,\ \ 1\le i< i' \leq g \label{H0ii}
\end{align} 
or we may be interested in simultaneous tests against one-sided alternatives 
\begin{equation} H_{1,ii'}:\ \ \mu_i-\mu_{i'}> 0\ \  1\leq i< i'\leq g \label{H1sided} \end{equation}
in such a way that the family of tests controls the FWER in strong sense.

Let $\hat{\bmbeta}$ be an estimator of $\bmbeta$ based on the reduced model $Y_{ij} = \bmx^\textsc{t} _{ij}\bmbeta+\varepsilon_{ij}$ and let $Y_{ij}(\hat{\bmbeta})=Y_{ij}-(\x1ij\hat{\beta_1}+\ldots+\xpij\hat{\beta_p})$, $i=1,\ldots, g; j=1,\ldots,n_i$ denote the aligned observations (reduced model residuals). Let $\Ripij(\hat{\bmbeta})$ denote the rank of $Y_{ij}(\hat{\bmbeta})$ among $Y_{i1}(\hat{\bmbeta}),\ldots,Y_{in_i}(\hat{\bmbeta});\ Y_{i'1}(\hat{\bmbeta}),\ldots,Y_{i'n_i'}(\hat{\bmbeta})$, the aligned observations of the $i-{th}$ and $i'-{th}$ samples $i<i'$.

  To formulate the rank statistics based on pairwise rankings of the aligned observations, we let $\bmY_i=(Y_{i1}, \ldots, Y_{in_i})^\textsc{t}$, $\bmY_{ii'}=(\bmY_i^\textsc{t}, \bmY_{i'}^\textsc{t})^\textsc{t}$,

\begin{equation}
\X1ii=\begin{bmatrix}\label{X1iip}
X^\textsc{t}_i, 
X^\textsc{t}_{i'}
\end{bmatrix} ^\textsc{t}
\end{equation}

\noindent where $X_i$ is the $n_i\times p$ matrix of covariates for the $i-th$ sample,\ $i=1, \cdots,g$, and

\begin{equation}
\Xii=[\bmx_{0,ii'},\X1ii]\label{Xiip}
\end{equation}

\noindent where  \begin{equation}\label{x0iip} \bmx_{0,ii'}=(n_i+n_{i'})^{-1}(n_{i'}\mathbf{1}^\textsc{t}_{n_i}, -n_i\mathbf{1}^\textsc{t}_{n_{i'}})^\textsc{t},\ \ 1\leq i<i'\leq g\end{equation} where $\mathbf{1}_{n}$ is an n-vector of 1's.
Then following \eref{model} we define the  linear model   for observations in the $i-th$ and $i'-th$ samples 

\begin{equation} \label{lmiip} \bmY_{ii'}=\bmx_{0,ii'}\mu_{ii'}+X_{1,ii'}\bmbeta + \bmvep_{ii'},\ \ 1\leq i<i'\leq g\end{equation} where $\mu_{ii'}=\mu_i-\mu_{i'}$. Without any loss of generality we assume that $\X1ii$ is centered, i.e. $\mathbf{1}^\textsc{t}_{n_i+n_{i'}}\X1ii = \mathbf{0}$ and hence $\Xii$ in \eref{Xiip} is also centered.

Let \begin{equation}\bmRipi(\hat{\bmbeta})=(\Ripi1(\hat{\bmbeta}),\ldots,\Ripini(\hat{\bmbeta});\Riip1(\hat{\bmbeta}),\ldots,\Riipnip(\hat{\bmbeta}))^\textsc{t} \label{Riip} 
\end{equation} 

\noindent be the vector of two-sample aligned ranks and for some integer $N$, let $a_N(k)$ denote scores that are generated by a square integrable function  $\phi$ as follows
\begin{equation}
a_N(k)=\phi [k/(N+1)],\ k=1,\cdots,N \label{score}
 \end{equation}
and define
\begin{equation}\label{sigma2phi}
\sigma_\phi^2=\int_0^1{(\phi(u)-\bar{\phi})^2}du 
\end{equation}
where $\bar{\phi}=\int^1_0\phi(u)du$. 

Although the rank-scores defined above and all statistics defined in the sequel depend on the sample size, for simplicity in notation, we suppress such dependence. In addition, we consider the scores to be centered, i.e. $a_c(k) = a(k) - \bar{a},\ \ k=1, \cdots,N$\ where $\bar{a}=(1/N)\sum_k a(k)$. Popular choices of rank-scores are the uniform or Wilcoxon scores 
\[a(k) = k/(N+1),\ \ k=1, \cdots,N\] and normal or van der Waarden scores \[a(k) = \Phi^{-1}[ k/(N+1)],\ \ k=1, \cdots,N\] where $\Phi^{-1}(\cdot)$\ is the inverse cdf of the standard normal distribution.

To test each hypothesis $H_{0,ii'}:\ \ \mu_i-\mu_{i'}=0, \ \ 1\le i< i' \leq g$ in relation to the two-sample linear model \eref{lmiip},  we use the aligned rank transform test statistic (Mansouri\cite{mansouri15})

\begin{equation}
T[\bmR_{ii'}(\hat{\bmbeta})]  =  \{\hat{\sigma}^2_{\phi, w}{\bmx^\textsc{T}_{0,ii'}}(I-H_{ii'})\bmx_{0,ii'}\}^{-1/2}\bmx^\textsc{T}_{0,ii'}(I-H_{ii'})a_c[\bmRipi(\hat{\bmbeta})]\label{Tiip}
\end{equation} 

\noindent
where $\hat{\sigma}^2_{\phi, w}$ defined in \eref{sigma2phiw} below, is a consistent estimator of $\sigma^2_\phi$ in \eref{sigma2phi}  and 

\begin{equation}H_{ii'} =X_{1, ii'} (X^\textsc{t}_{1, ii'}X_{1, ii'})^{-1}X^\textsc{t}_{1, ii'}\label{Hiip}\end{equation} is the hat matrix for the two-sample linear model \eref{lmiip}. It is instructive to note that $T_{ii'}[R(\hat{\bmbeta})]$ is the aligned rank transform analogue of the studentized least-squares estimate of $\mu_{ii'} = \mu_i-\mu_{i'}$ in the linear model \eref{lmiip}.  Straightforward calculations show that \eref{Tiip} can be written as 

\begin{equation} T[\bmR_{ii'}(\hat{\bmbeta})] =(\bar{a}_{i\cdot, ii'(adj)} -\bar{a}_{i'\cdot, ii'(adj)})/{SE_w(\bar{a}_{i\cdot, ii'(adj)} -\bar{a}_{i'\cdot, ii'(adj)})}\label{Tiip2} \end{equation} 
\noindent
where the adjusted mean rank-scores are defined as

 \[\bar{a}_{i\cdot, ii'(adj)}=\bar{a}_{i\cdot, ii'} -\bar{a}_{\cdot\cdot, ii'} - \bar{\bmx}^\textsc{t}_{i} (X^\textsc{t}_{1,ii'}X_{1,ii'})^{-1}X^\textsc{t}_{1,ii'} a_c[\bmRipi(\hat{\bmbeta})] \]

\begin{equation}\begin{aligned}SE( \bar{a}_{i\cdot, ii'(adj)} -\bar{a}_{i'\cdot, ii'(adj)})= & \{\hat{\sigma}^2_{\phi, w} (\frac{1}{n_i}+ \frac{1}{n_{i'}})\\ &[1- (\frac{1}{n_i}+ \frac{1}{n_{i'}})^{-1}(\bar{\bmx}^\textsc{t}_i,\ -\bar{\bmx}^\textsc{t}_{i'})(X^\textsc{t}_{1,ii'}X_{1,ii'})^{-1}(\bar{\bmx}^\textsc{t}_{i}, -\bar{\bmx}^\textsc{t}_{i'})^\textsc{t}]\}^{1/2}\nonumber \end{aligned} \end{equation}

\noindent
$\bar{a}_{i\cdot, ii'}$ is the mean rank-score for the $i-th$ sample when the $i-th$ and $i'-th$ samples are being compared, i.e.  $\bar{a}_{i\cdot, ii'}=\frac{1}{n_i}\sum_{j=1}^{n_i}a(R_{ij,ii'}),\ \ \bar{a}_{\cdot\cdot, ii'}=\frac{1}{n_i+n_{i'}}[\sum_{j=1}^{n_i}a(R_{ij,ii'})+\sum_{j'=1}^{n_{i'}}a(R_{i'j',ii'})], \mbox{and}\  \ \bar{\bmx}_i = \frac{1}{n_i}\mathbf{1}^\textsc{t}_{n_i} X_i$ is the vector of means of the  covariate matrix corresponding to the $i-th$ group, $\ \  i=1, \cdots,g$. 
We note that the numerator is difference of the adjusted means of the rank-scores of the $i-{th}$ and $i'-{th}$ samples and the second term inside the brackets in the denominator represents the amount of reduction in the variance due to the covariates. 
Define 

\begin{equation}\hat{\sigma}^2_{\phi, ii'} =  (n_i+n_{i'}-p-2)^{-1} a^\textsc{t}_c[\bmRipi(\hat{\bmbeta})][I -\Xii(X^\textsc{t}_{ii'}\Xii)^{-1}X^\textsc{t}_{ii'}] a_c[\bmRipi(\hat{\bmbeta})]\label{sigma2phiiip}\end{equation}
where $\Xii$ is defined in \eref{Xiip} and let

\begin{equation}\hat{\sigma}^2_{\phi,w} = \sum_{i<i'} w_{ii'} \hat{\sigma}^2_{\phi, ii'} \label{sigma2phiw}\end{equation} where

\[w_{ii'} = (n_i + n_{i'} -p-2)/\sum_{i<i'}(n_i + n_{i'} -p-2),\ \ 1\leq i<i'\leq g\]

The estimate defined in \eref{sigma2phiiip} is the mean-square-error (MSE) based on aligned ranks for the two-sample linear model \eref{lmiip}.  Under a sequence of contiguous alternatives to \eref{H0ii}, it follows from Mansouri\cite{mansouri99} that 

\[\hat{\sigma}^2_{\phi, ii'} \xrightarrow[]{P} \sigma^2_\phi,\ \mbox{as}\ \min_{1\leq i\leq g}n_i\rightarrow \infty\]

In addition, under a sequence of contiguous alternatives to the global null hypothesis, it follows that 

\[\hat{\sigma}^2_{\phi, w} \xrightarrow[]{P} \sigma^2_\phi,\ \mbox{as}\ \min_{1\leq i\leq g}n_i\rightarrow \infty\]

For models other than one-way layouts, the tests based on the aligned rank statistics \eref{Tiip2} are not distribution-free since the aligned observations are neither independent nor, in most cases, identically distributed and hence their ranks are not uniformly distributed. Even for one-way layouts where the tests are distribution-free, there are limitations on the  availability of tables of exact quantiles or tail probabilities. All existing tables for pairwise comparisons in one-way layouts are for comparing three groups. In most cases there is no choice but to use approximation methods of calculating  quantiles and p-values.

To conduct simultaneous tests for the pairwise comparisons of the treatment effects as formulated in \eref{H0ii}, each hypothesis $H_{0, ii'}: \mu_i-\mu_{i'} = 0$\ is rejected in favor of the two-sided alternative\ $H_{0, ii'}: \mu_i-\mu_{i'} \neq 0$, if 

\begin{equation}|T(\bmR_{ii'})|> q_\alpha \label{2sidedtest}\end{equation} where $q_\alpha$\ is the upper $\alpha$-th quantile of the distribution of 

\[Q_(\bmR) = \max_{1\leq i<i'\leq g} |T(\bmR_{ii'})|\] under the global null hypothesis $H_0: \mu_1=\cdots=\mu_g$.

To test against one-sided alternatives, $H_{0.ii'}$ is rejected if 

\begin{equation}T(\bmR_{ii'}) >q_{1, \alpha}\label{1sidedtest}\end{equation} where $q_{1, \alpha}$ is the upper $\alpha$-th quantile of the sampling distribution of 

\[Q_(\bmR) = \max_{1\leq i<i'\leq g} T(\bmR_{ii'})\] under the global null hypothesis $H_0: \mu_1=\cdots=\mu_g$.

%%%%%%%%%%%%%%%%%%%%%%%%%%%%%%%%%%%%%%%%%%%%%%%%%%%%%%%%%%%%%%%%%
%%%%%%%%%%%%%%%%%%%%%%%%%%%%%%%%%%%%%%%%%%%%%%%%%%%%%%%%%%%%%%%%%

\subsection{Treatments versus Control}

In treatment vs. control comparisons assuming that group g represents the control group, we are either interested in testing against two-sided alternatives 

\begin{align} H_{0,ii'}:&\ \ \mu_i- \mu_g =0 \nonumber\\ H_{1,ii'}:&\ \  \mu_i- \mu_g \neq 0, \ \ 1\leq i \leq g-1 \nonumber\end{align} or one-sided alternatives

\[H_{1, ii'}:\ \ \mu_i- \mu_g > 0, \ \ 1\leq i \leq g-1 \] 

\noindent It is noted that the problem of treatments vs. control is formulated as a special case of the all pairs comparisons defined in \eref{Tiip2}. The test statistic is  

\begin{equation} T^*(\bmR_{ig}) =(\bar{a}_{i\cdot, ig(adj)} -\bar{a}_{g\cdot, ig(adj)})/{SE(\bar{a}_{i\cdot, ig(adj)} -\bar{a}_{g\cdot, ig(adj)})},\ \ i=1, \cdots,g-1\label{TvC} \end{equation} 
\noindent
where \[\bar{a}_{i\cdot, ig(adj)}=\bar{a}_{i\cdot, ig}- \bar{a}_{\cdot\cdot, ig}-\bar{\bmx}^\textsc{t}_{i} (X^\textsc{t}_{1,ig}X_{1,ig})^{-1}X^\textsc{t}_{1,ig} a_c(\bmR_{ig}) \]

\begin{equation}\begin{aligned}SE( \bar{a}_{i\cdot, ig(adj)} -\bar{a}_{g\cdot, ig(adj)})= &\{\hat{\sigma}^2_{\phi, w} (\frac{1}{n_i}+ \frac{1}{n_{g}})\\&[1- (\frac{1}{n_i}+ \frac{1}{n_{g}})^{-1}(\bar{\bmx}_i -\bar{\bmx}_{g})^\textsc{t}(X^\textsc{t}_{1,ig}X_{1,ig})^{-1}(\bar{\bmx}_{i} -\bar{\bmx}_{g})]\}^{1/2}\nonumber \end{aligned} \end{equation}

 \[\hat{\sigma}^2_{\phi,wg} = \sum_{i=1}^{g-1} w_{ig} \hat{\sigma}^2_{\phi, ig} \]

\[\hat{\sigma}^2_{\phi, ig} =  (n_i+n_{g}-p-2)^{-1} a^\textsc{t}_c[(\bmR_{ig})][I -X_{ig}(X^\textsc{t}_{ig}X_{ig})^{-1}X^\textsc{t}_{ig}] a_c[(\bmR_{ig})]\]
and 
\[w_{ig} = (n_i + n_{g} -p-2)/\sum_{i=1}^{g-1}(n_i + n_g -p-2),\ \ 1\leq i\leq g-1\]

Then $H_{0,ig}: \mu_i- \mu_g =0$ is rejected in favor of the two-sided alternative if 

\[|T^*(\bmR_{ig})| > q^*_\alpha\] where $q^*_\alpha$ is the upper $\alpha$-th quantile of the sampling distribution of 

\[ Q^*(\bmR) = \max_{1\leq i\leq g-1}|T^*(\bmR_{ig})|\]

\noindent In addition, $H_{0,ii'}$ is rejected in favor of the one-sided alternative if

\[T^*(\bmR_{ig}) > q^*_{1,\alpha}\] where $q^*_{1,\alpha}$ is the upper $\alpha$-th quantile of the sampling distribution of 

\[ Q^*_1(\bmR) = \max_{1\leq i\leq g-1}T^*(\bmR_{ig})\]

%%%%%%%%%%%%%%%%%%%%%%%%%%%%%%%%%%%%%%%%%%%%%%%%%%%%%%%%%%%%%%%%%
%%%%%%%%%%%%%%%%%%%%%%%%%%%%%%%%%%%%%%%%%%%%%%%%%%%%%%%%%%%%%%%%%

\subsection{One-Way Layouts}

When no covariates are involved, the test statistics for pairwise comparisons are the studentized difference of the mean rank-scores of the i-th and i'-th samples given by

\begin{equation} 
T(\bmR_{ii'}) =(\bar{a}_{i\cdot, ii'} -\bar{a}_{i'\cdot, ii'})/{\{\hat{\sigma}^2_{\phi,w} (\frac{1}{n_i}+ \frac{1}{n_{i'}})\}^{1/2}},\ \ 1\leq i<i'\leq g\label{twosamp}
\end{equation}

where following \eref{sigma2phiiip} and \eref{sigma2phiw} as specialized to the one-way layouts, we have 

\begin{equation*}\hat{\sigma}^2_{\phi,w} = \sum_{i<i'} w_{ii'} \hat{\sigma}^2_{\phi, ii'} \end{equation*} 

\begin{equation*}\hat{\sigma}^2_{\phi, ii'} =  (n_i+n_{i'}-2)^{-1} [\sum_{j=1}^{n_i} \{a(R_{ij,ii'})-\bar{a}_{i\cdot, ii'}\}^2 +\sum_{j'=1}^{n_{i'}} \{a(R_{i'j',ii'})-\bar{a}_{i'\cdot, ii'}\} ^2] \end{equation*}

and

\[w_{ii'} = (n_i + n_{i'} -2)/\sum_{i<i'}(n_i + n_{i'} -2),\ \ 1\leq i<i'\leq g\]

The class of statistics in \eref{twosamp} present a generalization of the Steel-Dwass statistics which are obtained if the Wilcoxon scores \[ a(k) = {k}/(n_i+n_{i'}+1), \ \ k=1, \cdots,n_i+n_{i'}\] are used and the rank-scale parameter $\sigma^2_\phi$ is estimated by 

 \[\hat{\sigma}^2_{\phi, ii'} = {\sum_{k=1}^{n_i+n_{i'}} (a(k) - \bar{a})^2}/(n_i+n_{i'}-1)\] 
 see H\`{a}jek and \v{S}id\'{a}k\cite{hajek} for which the test statistic in \eref{twosamp} reduces to the standardized Wilcoxon statistic for comparing the $i-th$ and $i'-th$, Hochberg and Tamhane \cite{hochberg} and Critchlow and Fligner \cite{critchlow},

 \begin{equation} T(\bmR_{ii'}) = (\bar{R}_{i\cdot, ii'} - \bar{R}_{i'\cdot, ii'})/\{\frac{(n_i + n_{i'})(n_i + n_{i'} +1)}{12}(\frac{1}{n_i}+ \frac{1}{n_{i'}})\}^{1/2}\nonumber 
 \end{equation}

The class of statistics for the one-way layouts given in \eref{twosamp} are nonparametric distribution-free over the class continuous cdf's $F_i(x)= F(x) ,\ \ i=1, \cdots,g$, and hence an exact test two-sided (one-sided) that is based on the sampling distribution of the statistic $\max_{1\leq i<i'\leq g} |T(\bmR_{ii'})|$, ($\max_{1\leq i<i'\leq g} T(\bmR_{ii'})$), can be carried out that controls the familywise error rate exactly and in strong sense. However, as mentioned previously, exact tests are difficult to use and hence one usually resorts to conducting approximate tests.

For comparing treatments versus control the test statistics are given by

\begin{equation} 
T(\bmR_{ig}) =(\bar{a}_{i\cdot, ig} -\bar{a}_{g\cdot, ig})/\{\hat{\sigma}^2_{\phi,wg} (\frac{1}{n_i}+ \frac{1}{n_g})\}^{1/2},\ \  i=1, \cdots, g-1\label{twosampTvC}
\end{equation}

where

\[\hat{\sigma}^2_{\phi,wg} = \sum_{i=1}^{g-1} w_{ig} \hat{\sigma}^2_{\phi, ig} \]

\[\hat{\sigma}^2_{\phi, ig} =  (n_i+n_{g}-2)^{-1} [\sum_{j=1}^{n_i} \{a(R_{ij,ig})-\bar{a}_{i\cdot, ig}\}^2 +\sum_{j'=1}^{n_g} \{a(R_{gj',ig})-\bar{a}_{g\cdot, ig}\} ^2]\]
and 
\[w_{ig} = (n_i + n_{g} -2)/\sum_{i=1}^{g-1}(n_i + n_g -2),\ \ 1\leq i\leq g-1\]

The classes of statistics for the one-way layouts given in  \eref{twosampTvC} are nonparametric distribution-free over the class continuous cdf's $F_i(x) = F_g(x) ,\ \ i=1, \cdots,g-1$, and hence an exact two-sided (one-sided) test that is based on the sampling distribution of the statistic $\max_{1\leq i<i'\leq g} |T(\bmR_{ig})|$, ($\max_{1\leq i<i'\leq g} T(\bmR_{ig})$), can be carried out that controls the familywise error rate exactly and in strong sense.

%%%%%%%%%%%%%%%%%%%%%%%%%%%%%%%%%%%%%%%%%%%%%%%%%%%%%%%%%%%%%%%%%
%%%%%%%%%%%%%%%%%%%%%%%%%%%%%%%%%%%%%%%%%%%%%%%%%%%%%%%%%%%%%%%%%

 \section{Asymptotic Distribution of the Test Statistic}

In this section, the asymptotic distribution of the test statistic is derived. Define the vector of test statistics

\begin{equation}
\bmT(\bmR) = (T(\bmR_{12}), \cdots,T(\bmR_{1g}); T(\bmR_{23}), \cdots,T(\bmR_{2g}); \cdots; T_{g-1,g}(\bmR_{g-1,g}))^\textsc{t}
\label{T}
\end{equation}
where $T(\bmR_{ii'}), \ 1\leq i<i'\leq g$\ is defined in \eref{Tiip2}.  It is assumed that the error distribution function $F(\cdot)$ has
a density $f(\cdot)$ which is absolutely continuous and satisfies
\begin{equation}
\label{fisherInf} I(F)=\int^\infty_{-\infty}[f'(x)/f(x)]^2dF(x)<\infty
\end{equation}
where $f'(\cdot)$ is the derivative of $f(\cdot)$. Assume that

\begin{equation}
\lim_{N \longrightarrow \infty} {N^{-1}}X^\textsc{t}_{i} X_{i} = V_{i},\ 1\leq i'\leq g
\label{Vi}
\end{equation}
where $X_i$ is the matrix of covariate for the $i-th$ group, $V_{i'}$ is a positive definite matrix and assume that

\begin{equation} \label{lambdai} \lim_{N \longrightarrow \infty}({n_i}/{N}) = \lambda_i, \ 0\leq \lambda_i\leq 1, and \ \lambda_1 + \cdots+\lambda_g = 1\end{equation}

\noindent We further assume that
\begin{equation}
\lim_{N \longrightarrow \infty} \max_{1\leq j\leq n_i + n_{i'}} \bmx^\textsc{t}_{j, ii'}(X^\textsc{t}_{ii'}X_{ii'})^{-1}\bmx_{j,ii'} =0, \ \ 
\label{Noether}
\end{equation}

\noindent where $\bmx_{j,ii'}$ is the $j-th$ row of  $X_{ii'}$\  defined in \eref{Xiip}. Finally, we assume that $\hat{\bmbeta}$ is bounded in probability, i.e.

\begin{equation}
\sqrt{N}||\hat{\bmbeta}-\bmbeta||=O_p(1),\ \ as\ N\rightarrow\infty\label{betahat}
\end{equation}
where $N=\sum_in_i$ and $||\cdot||$ denotes the Euclidean norm.

\noindent Define
\begin{equation}
\mathbf{c}^\textsc{t}_{i,ii'}=(\frac{1}{n_i} + \frac{1}{n_{i'}})^{-1} \frac{1}{n_i}\bm1^\textsc{t}_{n_i} - \bmx^\textsc{t}_{0,ii'}X_{1,ii'} (X^\textsc{t}_{1,ii'}X_{1,ii'})^{-1}X^\textsc{t}_i
\label{ciiip}
\end{equation}

\noindent and 

\begin{equation}
\mathbf{c}^\textsc{t}_{i',ii'}=-(\frac{1}{n_i} + \frac{1}{n_{i'}})^{-1} \frac{1}{n_{i'}}\bm1^\textsc{t}_{n_{i'}} - \bmx^\textsc{t}_{0,ii'}X_{1,ii'} (X^\textsc{t}_{1,ii'}X_{1,ii'})^{-1}X^\textsc{t}_{i'}
\label{cipiip}
\end{equation}

\noindent We note that 

\begin{equation}\mathbf{c}^\textsc{t}_{ii'} =(\mathbf{c}^\textsc{t}_{i,ii'}\ ,\ \mathbf{c}^\textsc{t}_{i',ii'})= \bmx^\textsc{t}_{0,ii'} (I - H_{ii'}) \label{ciip}\end{equation}

\newtheorem{cor}[theorem]{Corolary}

\begin{theorem}\label{th:1}
Assume that \eref{fisherInf} - \eref{betahat}  hold, then under $H_{0, ii'}: \mu_i =\mu_{i'},\ 1\leq i<i'\leq g$,\ the statistic  $T_{ii'}$ defined in \eref{Tiip} has an asymptotic (as $N\longrightarrow \infty$)\ standard normal distribution.
\end{theorem}

\begin{theorem}\label{th:2}
Assume that the conditions of Theorem \ref{th:1}  hold, then under $H_0: \mu_1 = \cdots = \mu_g$,\ the vector-valued statistic  $\bmT$ defined in \eref{T} has a multivariate normal distribution with mean $\mathbf{0}$ and correlation matrix $C$ whose elements are given by  
	
\begin{equation}
	\rho_{ii', rr'} =\left\{\begin{array}{lll} 1 & & i=r,\ i'=r'\\ & & \\ \Lim{N \rightarrow \infty} \mathbf{c}^\textsc{t}_{i,ii'}\mathbf{c}_{i,ir'}/\sqrt{(\mathbf{c}^\textsc{t}_{ii'}\mathbf{c}_{ii'})(\mathbf{c}^\textsc{t}_{ir'}\mathbf{c}_{ir'})} & & i=r,\ i'\neq r' \nonumber \\
	& &\nonumber \\
\Lim{N \rightarrow \infty} \mathbf{c}^\textsc{t}_{i',ii'}\mathbf{c}_{i',ri'}/\sqrt{(\mathbf{c}^\textsc{t}_{ii'}\mathbf{c}_{ii'})(\mathbf{c}^\textsc{t}_{ri'}\mathbf{c}_{ri'})} & &  i\neq r,\ i'=r'\  \nonumber \\ & & \\
	\Lim{N \rightarrow \infty} \mathbf{c}^\textsc{t}_{i,ii'}\mathbf{c}_{i,ri}/\sqrt{(\mathbf{c}^\textsc{t}_{ii'}\mathbf{c}_{ii'})(\mathbf{c}^\textsc{t}_{ri}\mathbf{c}_{ri})} & & i=r',\ i'\neq r \nonumber \\
	& & \nonumber\\
	\Lim{N \rightarrow \infty} \mathbf{c}^\textsc{t}_{i',ii'}\mathbf{c}_{i',i'r'}/\sqrt{(\mathbf{c}^\textsc{t}_{ii'}\mathbf{c}_{ii'})(\mathbf{c}^\textsc{t}_{i'r'}\mathbf{c}_{i'r'})} & & i\neq r',\ i'= r \nonumber \\
	& & \nonumber\\
	0 & & \mbox{otherwise} 
\end{array} \label{corr}	\right.	 \end{equation}
where $\mathbf{c}_{i,ii'}, \mathbf{c}_{i',ii'}$\ and $\mathbf{c}_{ii'}$\ are given by \eref{ciiip}-\eref{ciip}.
\end{theorem}

\begin{cor}
Following Theorems \ref{th:1} and \ref{th:2}, the tests defined by \eref{2sidedtest} and \eref{1sidedtest} asymptotically (as $N\rightarrow \infty$) control the familywise error rate in strong sense.
\end{cor}

Note that we can write the  limiting correlations in terms of the limiting values \eref{Vi} and \eref{lambdai}, but the terms are too involved and do not contribute to simplicity in notation. In addition, in practice we use the estimated values that involve terms in the arguments of limit operator.
Proofs of the above theorems are provided in the Appendix.

\begin{cor}
Assume that the conditions of Theorem \ref{th:1}  hold, then under $H_0: \mu_1 = \cdots = \mu_g$,\ the vector of treatment vs. control statistics  $\bmT^*(\bmR) = (T^*(\bmR_{1g}), \cdots,T^*(\bmR_{g-1,g}))^\textsc{t}$ defined in \eref{TvC} has a multivariate normal distribution with mean $\mathbf{0}$ and correlation matrix $C^*$ whose elements are given by  

\begin{equation}
\rho_{ig, rg} =\left\{\begin{array}{lll} 1 & & i= r (= 1, \cdots,g-1)\\ & & \\ 
\Lim{N \rightarrow \infty} \mathbf{c}^\textsc{t}_{g,ig}\mathbf{c}_{g,rg}/\sqrt{(\mathbf{c}^\textsc{t}_{ig}\mathbf{c}_{ig})(\mathbf{c}^\textsc{t}_{rg}\mathbf{c}_{rg})} & &  i\neq r \nonumber 
\end{array} \label{corr}	\right.	 \end{equation}
\end{cor}

%%%%%%%%%%%%%%%%%%%%%%%%%%%%%%%%%%%%%%%%%%%%%%%%%%%%%%%%%%%%%%%%%
%%%%%%%%%%%%%%%%%%%%%%%%%%%%%%%%%%%%%%%%%%%%%%%%%%%%%%%%%%%%%%%%%

\section{Simulation Study}

It was noted in section 2 that asymptotic approximation is the main method of conducting multiple comparisons based on rank statistics. Following the results in section 3, one is inclined to use the distribution of maximum modulus for two-sided tests, and maximum for one-sided tests, of a vector of multivariate normally distributed (MVN) random variable as the sampling distribution of the pairwise test statistic given in \eref{2sidedtest} and \eref{1sidedtest}, respectively. However, simulation study of Mansouri\cite{mansouri15} concerning the aligned rank transformation method based on overall ranking of all observations indicate that using the relevant sampling distribution based on multivariate t-distribution (MVT) results in a test that has the robustness of validity property by controlling the FWER at a prescribed nominal level. 

To investigate the robustness of validity property of the test based on pairwise rankings, we simulate a balanced one-way analysis of covariance model involving six treatments and one covariate. Specifically, we consider the model $Y_{ij}=\mu_i+\beta X+\epsilon_{ij}$,\ $i=1,\ldots, 6, j=1,\ldots, n$, where $\beta =5$. The covariate $X$ is generated from a standard normal distribution. The error terms $\epsilon_{ij}$ are generated from the following distributions:
 
\begin{itemize}
\item[1.] Standard normal distribution. 

\item[2.] Standard lognormal distribution. 

\item[3.] Cauchy distribution. 

\item[4.] Heteroscedastic normal (H-N) distributions with mean zero and variances\\ $(\sigma_1^2, \sigma_2^2, \sigma_3^2, \sigma_4^2, \sigma_5^2, \sigma_6^2)=(1,1,2,2,4,4)$.

\end{itemize}

 To calculate empirical FWER , we set $\mu_i=2,\ i=1,\ldots, 6$ and to calculate empirical power, we set $(\mu_1, \mu_2, \ldots, \mu_6)=(1,1,2,2,4,4)$. Under each setting, the FWER and power are obtained based on $10,000$ simulations. For each simulation, independent $\epsilon$'s  are generated while same values of $X$ are repeated.

  We consider four tests  for comparison. 
  \begin{itemize}
  	\item[1.] Test based on statistics defined in \eref{Tiip2} which will be referred to  as PWR\_W, statistic. These statistics are based on pairwise rankings using the weighted estimate $\hat{\sigma}^2_{\phi, w}$ in \eref{sigma2phiw} of the scale parameter $\sigma^2_\phi$. 
  	
  		\item[2.]  Test based on the above statistics except that the two-sample estimates $\hat{\sigma}^2_{\phi, ii'}$ defined in \eref{sigma2phiiip} are used instead of $\hat{\sigma}^2_{\phi, w}$. This test will be referred to as PWR.
  		
  		\item[3.]  The aligned rank-transformation (ART) test of Mansouri \cite{mansouri15}. 
  		
  			\item[4.]  The least squares (LS) test.
  			
  	\end{itemize}
  
  We investigate the robustness of validity of PWR\_W and PWR using the upper quantile of maximum modulus of MVN as well as MVT with Sattertwaite approximation of the degrees of freedom. For ART and LS, we use quantiles of MVT, Mansouri\cite{mansouri15}. 
  
  Control of FWER as well as power comparisons are studied for models with $n=10$ and $n=15$ each for FWERs of $\alpha = 0.01$ and $\alpha=0.05$. These results are summarized in Tables 1-3. Table 1 summarizes the simulated FWER. It is clear that PWR test does not control the FWER at the nominal level. In fact in all cases the test is anti-conservative and over-estimates the nominal level. On the other hand, PWR\_W controls the FWER at the nominal level for both MVT and MVN approximation of the quantiles as compared with the 95\% tolerance intervals of ($0.01\pm 0.002$ and $0.05 \pm 0.004$). In most cases using MVN approximation of quantiles results in slightly less conservative test. Furthermore, PWR\_W is the only test that controls FWER for the heteroscedastic normal (H-N) distribution. Surprisingly, for the larger sample size where $n=15$ and $\alpha=0.05$, the estimated FWER is slightly higher than the upper tolerance limit, yet it is considerably lower than the estimated FWER for ART and LS.  ART performs well in all cases except for H-N distribution, under which the FWER is inflated. The LS test produces conservative tests for non-normal distributions and is anti-conservative for H-N distribution.

  Table 2 contains comparisons of the four methods with respect to minimal power,  the probability of rejecting at least one of the false null hypotheses. We note that power comparison involving PWR method is not meaningful since it does not control FWER at the nominal level. Same interpretation holds true for ART and LS for the case of H-N distribution. For cases where power comparison is appropriate, LS has low power as compared with PWR\_W and ART. PWR\_W generally has higher minimal power than ART.
  
  For proportional power, the average proportion of false null hypotheses that are rejected, the results are summarized in Table 3, both PWR\_W and ART have higher power as compared to LS method including for normally distributed data. For lognormal and Cauchy, there does not seem to be a distinct advantage for the competing rank tests.
    
\newpage
\begin{scriptsize}
	\begin{longtable}{clcccc}
		\caption{Familywise Error Rate \label{table: FWER}}\\
		
			\multicolumn{6}{c}{$n=10$\ $\alpha = 0.01$}\\ 	
			& & \multicolumn{4}{c}{Distribution}\\ \cline{3-6}\\		
		Critical Value	&	Test	&	Normal	&	Lognormal	&	Cauchy	&	H-N\footnote{Heteroscedastic normal}		\\
		\hline
		MVT\footnote{Quantile of max-modulus of multivariate t}	&	PWR\_W\footnote{Test based on pairwise rankings in \eref{Tiip2}}	&	0.005	&	0.005	&	0.006	&	0.008	\\
		MVN\footnote{Quantile of max-modulus of multivariate normal}		&	PWR\_W	&	0.006	&	0.006	&	0.007	&	0.009	\\
		MVT	&	PWR\footnote{Pairwise rankings statistics using two-sample scale estimates \eref{sigma2phiiip}}		&	0.018	&	0.017	&	0.013	&	0.024	\\
		MVN	&	PWR	&	0.054	&	0.056	&	0.047	&	0.064	\\	
		MVT	&		ART	&	0.009	&	0.008	&	0.009	&	0.019	\\
		MVT	&	LS	&	0.009	&	0.005	&	0.003	&	0.035	\\		
		\hline
		\multicolumn{6}{c}{$n=15$\ $\alpha = 0.01$}\\ 	
		\hline
		MVT	&	PWR\_W	&	0.007	&	0.008	&	0.009	&	0.011	\\
		MVN	&	PWR\_W	&	0.008	&	0.009	&	0.011	&	0.012	\\	
		MVT	&	PWR	&	0.017	&	0.016	&	0.010	&	0.022	\\
		MVN	&	PWR		&	0.034	&	0.034	&	0.024	&	0.045	\\
		
		MVT	&	ART		&	0.009	&	0.009	&	0.010	&	0.024	\\
		MVT	&	LS	&	0.009	&	0.006	&	0.003	&	0.036	\\
		
		\hline
		\multicolumn{6}{c}{$n=10$\ $\alpha = 0.05$}\\
	 \hline	
		MVT	&	PWR\_W	&	0.037	&	0.037	&	0.046	&	0.046	\\
		MVN	& 	PWR\_W	&	0.041	&	0.043	&	0.050	&	0.049	\\
		MVT	&	PWR		&	0.073	&	0.073	&	0.063	&	0.083	\\
		MVN	&	PWR		&	0.134	&	0.131	&	0.116	&	0.145	\\
		
		MVT	&	ART		&	0.048	&	0.048	&	0.050	&	0.069	\\
		MVT	&	LS	&	0.047	&	0.036	&	0.027	&	0.095	\\
		
		\hline
		\multicolumn{6}{c}{$n=15$\ $\alpha = 0.05$}\\ 	
		\hline
		MVT	&	PWR\_W	&	0.045	&	0.045	&	0.049	&	0.056	\\
		MVN	&	PWR\_W	&	0.046	&	0.046	&	0.050	&	0.058	\\
		MVT	&	PWR		&	0.065	&	0.064	&	0.050	&	0.078	\\
		MVN	&	PWR	&	0.096	&	0.097	&	0.081	&	0.112	\\
		
		MVT	&	ART		&	0.048	&	0.048	&	0.049	&	0.077	\\
		MVT	&	LS	&	0.048	&	0.039	&	0.024	&	0.099	\\
	
		\hline
	\end{longtable}
\end{scriptsize}

\newpage
\begin{scriptsize}
	\begin{longtable}{clcccc}
		\caption{Minimal Power \label{table: Min Power}}\\
		\hline
		\multicolumn{6}{c}{$n=10$\ $\alpha = 0.01$}\\ 	
	& & \multicolumn{4}{c}{Distribution}\\ \cline{3-6}\\		
	Critical Value	&	Test	&	Normal	&	Lognormal	&	Cauchy	&	H-N	\\
	\hline
		MVT	&		PWR\_W	&	1.000	&	0.960	&	0.477	&	0.257	\\
		MVN		&	PWR\_W	&	1.000	&	0.967	&	0.496	&	0.272	\\
		MVT		&		PWR	&	1.000	&	0.976	&	0.545	&	0.387	\\
		MVN		&	PWR	&	1.000	&	0.995	&	0.720	&	0.580	\\
		MVT		&	ART	&	1.000	&	0.984	&	0.457	&	0.407	\\
		MVT		&	LS	&	1.000	&	0.743	&	0.054	&	0.413	\\
		
			\multicolumn{6}{c}{$n=15$\ $\alpha = 0.01$}\\ 
		\hline
		MVT		&	PWR\_W	&	1.000	&	0.999	&	0.753	&	0.517	\\
		MVN		&	PWR\_W	&	1.000	&	1.000	&	0.764	&	0.534	\\
		MVT		&	PWR	&	1.000	&	1.000	&	0.741	&	0.588	\\
		MVN		&	PWR	&	1.000	&	1.000	&	0.836	&	0.716	\\
		MVT		&	ART	&	1.000	&	1.000	&	0.680	&	0.648	\\
		MVT		&	LS	&	1.000	&	0.882	&	0.057	&	0.656	\\
			
				\multicolumn{6}{c}{$n=10$\ $\alpha = 0.05$}\\
		\hline
		MVT		&	PWR\_W	&	1.000	&	0.996	&	0.732	&	0.543	\\
		MVN		&	PWR\_W	&	1.000	&	0.996	&	0.742	&	0.557	\\
		MVT		&	PWR	&	1.000	&	0.997	&	0.757	&	0.628	\\
		MVN		&	PWR	&	1.000	&	0.999	&	0.836	&	0.741	\\			
		MVT		& ART	&	1.000	&	0.998	&	0.672	&	0.643	\\			
		MVT		&	LS	&	1.000	&	0.868	&	0.141	&	0.642	\\
			
				\multicolumn{6}{c}{$n=15$\ $\alpha = 0.05$}\\
		\hline
			
		MVT		&	PWR\_W	&	1.000	&	1.000	&	0.903	&	0.779	\\
		MVN		&	PWR\_W	&	1.000	&	1.000	&	0.905	&	0.781	\\
		MVT		&	PWR	&	1.000	&	1.000	&	0.893	&	0.811	\\
		MVN		&PWR	&	1.000	&	1.000	&	0.924	&	0.867	\\		
		MVT		&ART	&	1.000	&	1.000	&	0.825	&	0.838	\\
		MVT		&	LS	&	1.000	&	0.945	&	0.146	&	0.847	\\
			
		\hline
	\end{longtable}
\end{scriptsize}

\newpage
\begin{scriptsize}
	\begin{longtable}{clcccc}
		\caption{Proportional Power \label{table: Pro Power}}\\
		\hline
		\multicolumn{6}{c}{$n=10$\ $\alpha = 0.01$}\\ 	
		& & \multicolumn{4}{c}{Distribution}\\ \cline{3-6}\\		
		Critical Value	&	Test	&	Normal	&	Lognormal	&	Cauchy	&	H-N	\\
		\hline		
		MVT	&	PWR\_W	&	0.659	&	0.442	&	0.086	&	0.047	\\
		MVN	&	PWR\_W	&	0.669	&	0.456	&	0.091	&	0.051	\\
		MVT	&	PWR	&	0.579	&	0.386	&	0.090	&	0.067	\\
		MVN	&	PWR	&	0.665	&	0.507	&	0.153	&	0.122	\\
		MVT	&	ART	&	0.650	&	0.475	&	0.099	&	0.079	\\
		MVT	&	LS	&	0.636	&	0.250	&	0.008	&	0.093	\\
				\multicolumn{6}{c}{$n=15$\ $\alpha = 0.01$}\\
		\hline	
		MVT	&	PWR\_W	&	0.789	&	0.672	&	0.176	&	0.113	\\
		MVN	&	PWR\_W	&	0.794	&	0.679	&	0.181	&	0.118	\\
		MVT	&	PWR	&	0.711	&	0.590	&	0.158	&	0.123	\\
		MVN	&	PWR\_W	&	0.750	&	0.659	&	0.211	&	0.172	\\
		MVT	&	ART	&	0.766	&	0.634	&	0.182	&	0.146	\\
		MVT	&	LS	&	0.736	&	0.357	&	0.008	&	0.170	\\
		\multicolumn{6}{c}{$n=10$\ $\alpha = 0.05$}\\
		\hline		
		MVT	&	PWR\_W	&	0.764	&	0.606	&	0.177	&	0.128	\\
		MVN	&	PWR\_W	&	0.769	&	0.614	&	0.182	&	0.133	\\
		MVT	&	PWR	&	0.686	&	0.536	&	0.171	&	0.140	\\
		MVN	&	PWR	&	0.735	&	0.608	&	0.220	&	0.192	\\
		MVT	&	ART	&	0.744	&	0.606	&	0.186	&	0.158	\\
		MVT	&	LS	&	0.722	&	0.363	&	0.023	&	0.176	\\
		\multicolumn{6}{c}{$n=15$\ $\alpha = 0.05$}\\
		\hline		
		MVT	&	PWR\_W	&	0.859	&	0.784	&	0.287	&	0.230	\\
		MVN	&	PWR\_W	&	0.861	&	0.786	&	0.289	&	0.232	\\
		MVT	&	PWR	&	0.786	&	0.715	&	0.263	&	0.228	\\
		MVN	&	PWR	&	0.812	&	0.752	&	0.303	&	0.272	\\
		MVT	&	ART 	&	0.838	&	0.735	&	0.290	&	0.248	\\
		MVT	&	LS	&	0.808	&	0.472	&	0.024	&	0.275	\\
				\hline
	\end{longtable}
\end{scriptsize}
\newpage

%%%%%%%%%%%%%%%%%%%%%%%%%%%%%%%%%%%%%%%%%%%%%%%%%%%%%%%%%%%%%%%%%%%%%%%%%%
\section{Real Data Analysis}
In a study to assess the short-term effects of sulfur dioxide(SO$_2$) exposure under various conditions, twenty-two young asthmatic volunteers were recruited, see~\cite{linn}. Their lung function (as defined by forced expiratory volume/fnorced vital capacity) and  baseline data regarding bronchial reactivity to SO$_2$ were measured at screening. The volunteers were stratified into three groups based on their lung function ($\le 74\%$, $75-84\%$, and $\ge 85\%$) with sample size 5, 12, and 5, respectively. It is of interest to compare the bronchial reactivity baseline means of each pair of the groups. 

This is an unbalanced one-way layout problem. We calculated PWR\_W test statistics based on the test defined in (\ref{twosamp}), and obtained the p-values using MVN and MVT. For comparison, the extended Steel-Dwass test given by \cite{critchlow} (SDCF) was also applied. The results are shown in Table \ref{real}.

\begin{table}[!htbp]
\caption{Pulmonary Disease Data Analysis Results.}
\begin{tabular}{ccccccc}\label{real}
\\
\hline
 	&	 \multicolumn{3}{c}{PWR\_W}&&\multicolumn{2}{c}{SDCF}\\							
	\cline{2-4} \cline{6-7}
\small{Comparison}	&	\small{stat.}	&	\small{adj. p-value(MVN)}	&	\small{adj. p-value(MVT)}	&&	\small{stat.}	&	\small{adj. p-value(MVN)}	\\
\hline
1-2	&	1.939	&	0.066	&	0.074	&	&	1.792 	&	0.090	\\
1-3	&	2.107	&	0.045	&	0.052	&	&	1.984	&	0.059	\\
2-3	&	1.141	&	0.274	&	0.280	&	&	 0.949	&	0.355	\\
\hline
\end{tabular}
\end{table}
It is noted that PWR\_W results in smaller p-values as compared with SDCF. It shows evidence of unequal bronchial reactivity baseline difference between groups with lung function $\le 74\%$ and $\ge 85\%$, at an adjusted p-values  $<0.05$.
%%%%%%%%%%%%%%%%%%%%%%%%%%%%%%%%%%%%%%%%%%%%%%%%%%%%%%%%%%%%%%%%%
%%%%%%%%%%%%%%%%%%%%%%%%%%%%%%%%%%%%%%%%%%%%%%%%%%%%%%%%%%%%%%%%%

\section{Concluding Remarks and Summary}

In this article we have extended the method of pairwise ranking for one-way layouts to the analysis of covariance model and factorial designs in general by combining the aligned ranking and pairwise ranking methods. The tests are distribution-free for one-way layouts and asymptotically distribution-free for higher order layouts. They control the familywise error rate in strong sense. Theses tests are developed for one-sided and two-sided comparisons as well as treatments vs. control problems. A weighted estimate  of rank-scale  is used to improve the small sample performance of the tests. This method should provide a better substitute for the existing generalized Steel-Dwass pairwise comparisons based on ranks for one-way layouts.

%%%%%%%%%%%%%%%%%%%%%%%%%%%%%%%%%%%%%%%%%%%%%%%%%%%%%%%%%%%%%%%%%%%%%%%%%%%%%%

\setcounter{section}{1}
\renewcommand{\theequation}{\Alph{section}\arabic{equation}}
\setcounter{equation}{0}
\section*{Appendix}

\noindent  \textit{Proof of Theorem 1}.  Define 
\begin{equation}\label{thetaiip}
\hat{\theta}_{ii'}(\hat{\bmbeta})=[\bmx^\textsc{t}_{0,ii'}(I-H_{ii'})\bmx_{0,ii'}]^{-1}\bmx^\textsc{t}_{0,ii'}(I-H_{ii'})a[\bmRipi(\hat{\bmbeta})]
\end{equation}
where \[H_{ii'}=X_{1,ii'}(X^\textsc{t}_{1,ii'}X_{1,ii'})^{-1}X^\textsc{t}_{1,ii'}\] is as defined in \eref{Hiip},  $\bmx_{0,ii'}$ is defined in \eref{x0iip} and $X_{1,ii'}$ is defined in \eref{X1iip}. Under the two-sample linear model \eref{lmiip} and under $H_{0, ii'}: \mu_i - \mu_{i'} = 0$, it follows from Theorem 3.1 of \cite{mansouri99} that

\begin{equation}
\sqrt{N}\hat{\theta}_{ii'}(\hat{\bmbeta})\xrightarrow[]{D} N(0,\sigma^2_{\phi}\sigma_{ii'}^{-1})),\ as\ N\rightarrow\infty\label{thetaNormal}
\end{equation}

\noindent where  $\sigma^2_{\phi}$ is defined in \eref{sigma2phi} and $\sigma_{ii'}$ is given by 
\begin{equation}\label{sigmaiip}
\sigma_{ii'}=\Lim{N\rightarrow\infty}N^{-1}\bmx^\textsc{t}_{0,ii'}(I - H_{ii'})\bmx_{0,ii'}
\end{equation}

\noindent Now consider $T[\bmR_{ii'}(\hat{\bmbeta})]$ as defined in \eref{Tiip} and it follows that 

\[
T[\bmR_{ii'}(\hat{\bmbeta})]  \approx  (\sigma^2_{\phi})^{-1/2}(\sigma_{ii'})^{1/2}N^{1/2}\hat{\theta}_{ii'}(\hat{\bmbeta})
\]

\noindent Hence it follows that under $H_{0,ii'}$ \[T[\bmR_{ii'}(\hat{\bmbeta})] \xrightarrow[]{D} N(0,\ 1)\,\, \mbox{as}\, N\rightarrow \infty, \,\, 1\leq i<i'\leq g \] This completes the proof.

%%%%%%%%%%%%%%%%%%%%%%%%%%%%%%%%%%%%%%%%%%%%%%%%%%%%%%%%%%%%%%%%%%%%%%%%%%%%%%%%%%%%%%%%%%%%%%

\begin{lemma}\label{th:A1}
Define \[\bmS^*(\bmbeta) = (S^*_{12}(\bmbeta), \cdots,S^*_{1g}(\bmbeta); \cdots; S^*_{g-1,g}(\bmbeta))^\textsc{t}\] where 

\begin{equation} S^*_{ii'}(\bmbeta)=\bmx^\textsc{t}_{0,ii'}(I-H_{ii'})  a[\bmRipi(\bmbeta)]\label{s*iip},\ \ 1\leq i<i'\leq g  \end{equation}

Then under the assumptions of Theorem \ref{th:1} \[ N^{-1/2}\bmS^*(\bmbeta) \xrightarrow{D} N(\mathbf{0}, \Sigma),\ \ \mbox{as}\ N\rightarrow \infty \]

\noindent where $\Sigma = ((\Lim{N\rightarrow\infty}   Cov(N^{-1/2}S^*_{ii'},N^{-1/2}S^*_{rr'})))$

\begin{align}
 \Lim{N\rightarrow\infty}  Cov(N^{-1/2}S^*_{ii'},N^{-1/2}S^*_{rr'})=  \begin{matrix}\left\{\begin{array}{ll}
 \sigma^2_{\phi}\Lim{N\rightarrow\infty}N^{-1}\bmc^\textsc{t}_{ii'}\bmc_{ii'} &i=r\ ,\ i'=r'\\ &\\
 \sigma^2_{\phi}\Lim{N\rightarrow\infty}N^{-1}\bmc^\textsc{t}_{i,ii'}\bmc_{i,ir'}&i=r\ ,\ i'\ne r'\\ &\\
 \sigma^2_{\phi}\Lim{N\rightarrow\infty}N^{-1}\bmc^\textsc{t}_{i',ii'}\bmc_{i',ri'}&i\ne r\ ,\ i'= r'\\ &\\
 \sigma^2_{\phi}\Lim{N\rightarrow\infty}N^{-1}\bmc^\textsc{t}_{i,ii'}\bmc_{i,ri}&i=r'\ ,\ i'\neq r\\ &\\
 \sigma^2_{\phi}\Lim{N\rightarrow\infty}N^{-1}\bmc^\textsc{t}_{i',ii'}\bmc_{i',i'r'}& i\neq r'\ ,\  i'= r\\ &\\ 0 & \mbox{otherwise}
 \end{array}\right.\end{matrix}\label{covar}
\end{align}

\noindent $\bmc_{i,ii'}$,\ $\bmc_{i',ii'}$,\ and $\bmc_{ii'}$ are vectors defined in \eref{ciiip}-\eref{ciip}, and  $\sigma^2_\phi$ is given by \eref{sigma2phi}
	
\end{lemma}	

\noindent {\textit Proof}: The distribution of $S^*_{ii'}(\bmbeta)$ under the assumption that
	\[F_{ij}(y)=P(Y_{ij}\le y)=F(y-\sum_{k=1}^px_{k,ij}\beta_k)\]
	is the same as the distribution of $S^*_{ii'}(\mathbf{0})$ under the assumption $F_{ij}(y)=F(y)$, $\forall ij$. Elements of $H_{ii'}$ defined in \eref{Hiip} are the leverage values of the matrix of covariates that can be expressed as
	
	\begin{align}
	H_{ii'}=&\X1ii(X^\textsc{t}_{1,ii'}\X1ii)^{-1}X^\textsc{t}_{1,ii'}\nonumber\\
	=&\begin{bmatrix}
	X_i\nonumber\\
	X_{i'}
	\end{bmatrix}(X^\textsc{t}_{1,ii'}\X1ii)^{-1}\begin{bmatrix}
	X^\textsc{t}_i X^\textsc{t}_{i'}
	\end{bmatrix} =(( h_{rj,r'j'})) 
	\end{align}
	
\noindent	where the matrix entries $h_{rj,r'j'} = \bmx^\textsc{t}_{rj}(X^\textsc{t}_{1,ii'}X_{1,ii'})^{-1}\bmx_{r'j'}$\ for $ r, r' = i, i', 1\leq i<i'\leq g$ and where $\bmx^\textsc{t}_{rk}$ is the $k-th$ row of $X_r$\ for $k=1, \cdots,n_r, \ r=1, \cdots,g$. Now the statistic $S^*_{ii'}$ can be written as
	
	\begin{align}
	S^*_{ii'}=&\bmx^\textsc{t}_{0.ii'}[I-H_{ii'}]a_c[\bmRipi]\nonumber\\
	=&(n_i+n_{i'})^{-1}[\sum_{j=1}^{n_i}\{n_{i'}(1-\sum_{k=1}^{n_i}h_{ik,ij})+n_i\sum_{j'=1}^{n_{i'}}h_{i'j',ij}\}a_c(R_{ij,ii'})\nonumber\\
	&\ \ -\sum_{j'=1}^{n_{i'}}\{n_{i'}\sum_{j=1}^{n_i}h_{ij,i'j'}+n_i(1-\sum_{k'=1}^{n_{i'}}h_{i'k',i'j'})\}a_c(R_{i'j',ii'})]\nonumber
	\end{align}
	where 
	\begin{equation}c_{ij,ii'}=(n_i+n_{i'})^{-1}\{n_{i'}(1-\sum_{k=1}^{n_i}h_{ik,ij})+n_i\sum_{j'=1}^{n_{i'}}h_{i'j',ij}\}\label{cijiip}\end{equation}
	
	\begin{equation}c_{i'j',ii'}=-(n_i+n_{i'})^{-1}\{n_{i'}\sum_{j=1}^{n_i}h_{ij,i'j'}+n_i(1-\sum_{k'=1}^{n_{i'}}h_{i'k',i'j'})\}\label{cipjpiip}\end{equation}
	Note that $c_{ij,ii'}$ and $c_{i'j',ii'}$ are elements of the vectors defined in \eref{ciiip} and \eref{cipiip}, respectively.
	
	Following the projection method of H\'{a}jek\cite{hajek68} Theorem 4.2 ,  the distribution of $S^*_{ii'}-E(S^*_{ii'})$ is the same as  distribution of $\sum_{r\in\{i,i'\}}\sum_{k=1}^{n_r}Z_{rk,ii'}$, where for  $r\in{i,i'}, k=1,\ldots,n_r  $
	\begin{equation}\label{ziip}
	Z_{rk,ii'}=(n_i+n_{i'})^{-1}\sum_{s\in\{i,i'\}}\sum_{l=1}^{n_s}(c_{sl,ii'}-c_{rk,ii'})\int{[u(y-Y_{rk})-F(y)]}\phi'[F(y)]dF(y)
	\end{equation}
	Since it can be shown that 
	\begin{equation}\nonumber
	\sum_{r\in\{i,i'\}}\sum_{k=1}^{n_r}c_{rk,ii'}=0
	\end{equation}
	%\[\sum_{k=1}^{n_i}c_{ik,ii'}+\sum_{k'=1}^{n_{i'}}d_{i'k',ii'}=0.\]
	it follows that  $Z_{rk,ii'}=-c_{rk,ii'}t(Y_{rk})$, where 
	\[t(Y_{rk})=\int{[u(y-Y_{rk})-F_0(y)]}\phi'[F_0(y)]dF_0(y),\ r= i,i';\ k=1,\ldots,n_r\]
	Furthermore, since $var(t(Y_{ij}))=\sigma_{\phi}^2$, $\forall i,j,$ %(see proof of Theorem 2.3 of Koziol and Reid (1977)), 
	it follows
	\begin{align}
	Var(Z_{rj,ii'})&=\sigma^2_{\phi}c_{rj,ii'}^2 & r=i,\ i'\nonumber\\ &&\nonumber\\ 
	Cov(Z_{ij,ii'},Z_{ij,ir'})&=\sigma^2_{\phi}c_{ij,ii'}c_{ij,ir'}& i<i'\neq r'\nonumber\\ &&\nonumber\\
	Cov(Z_{i'j',ii'},Z_{i'j',ri'})&=\sigma^2_{\phi}c_{i'j',ii'}c_{i'j',ri'}& i\neq r<i'\nonumber\\ &&\nonumber\\            
	Cov(Z_{i'j',ii'},Z_{i'j',i'r'})&=\sigma^2_{\phi}c_{i'j',ii'}c_{i'j',i'r'}& i<i'<r'\nonumber\\ &&\nonumber\\
	Cov(Z_{ij,ii'},Z_{ij,ri})&=\sigma^2_{\phi}c_{ij,ii'}c_{ij,ri}&r<i<i'\nonumber
	\end{align}
	
\noindent 	Hence for $1\leq i<i', r<r'\leq g$, the limiting covariance of $S^*_{ii'}=\sum_{j=1}^{n_i}Z_{ij,ii'}+\sum_{j'=1}^{n_{i'}}Z_{i'j',ii'}$ and $S^*_{rr'}=\sum_{j=1}^{n_r}Z_{rl,rr'}+\sum_{j'=1}^{n_{r'}}Z_{r'l',rr'}$ under $H_0$ is

\begin{align}
\Sigma =&	\Lim{N\rightarrow\infty}   Cov(N^{-1/2}S^*_{ii'},N^{-1/2}S^*_{rr'}) \nonumber\\ =&\Lim{N\rightarrow\infty}N^{-1}\{\sum_{j=1}^{n_i}\sum_{l=1}^{n_r}cov(Z_{ij,ii'},Z_{rl,rr'})
	+\sum_{j=1}^{n_i}\sum_{l'=1}^{n_{r'}}cov(Z_{ij,ii'},Z_{r'l',rr'})\nonumber\\ 
	&+\sum_{j'=1}^{n_{i'}}\sum_{l=1}^{n_r}cov(Z_{i'j',ii'},Z_{rl,rr'})
	+\sum_{j'=1}^{n_{i'}}\sum_{l'=1}^{n_{r'}}cov(Z_{i'j',ii'},Z_{r'l',rr'})\}\nonumber\\
	=&\begin{matrix}\left\{\begin{array}{ll}
	\sigma^2_{\phi}\Lim{N\rightarrow\infty}N^{-1}(\sum_{j=1}^{n_i}c_{ij,ii'}^2+\sum_{j'=1}^{n_i'}c_{i'j',ii'}^2) =\sigma^2_{\phi}\Lim{N\rightarrow\infty}N^{-1}\bmc^\textsc{t}_{ii'}\bmc_{ii'} &i=r\ ,\ i'=r'\\ &\\
	\sigma^2_{\phi}\Lim{N\rightarrow\infty}N^{-1}\sum_{j=1}^{n_i}c_{ij,ii'}c_{ij,ir'}=\sigma^2_{\phi}\Lim{N\rightarrow\infty}N^{-1}\bmc^\textsc{t}_{i,ii'}\bmc_{i,ir'}&i=r\ ,\ i'\ne r'\\ &\\
	\sigma^2_{\phi}\Lim{N\rightarrow\infty}N^{-1}\sum_{j'=1}^{n_{i'}}c_{i'j',ii'}c_{i'j',ri'}=\sigma^2_{\phi}\Lim{N\rightarrow\infty}N^{-1}\bmc^\textsc{t}_{i',ii'}\bmc_{i',ri'}&i\ne r\ ,\ i'= r'\\ &\\
	\sigma^2_{\phi}\Lim{N\rightarrow\infty}N^{-1}\sum_{j=1}^{n_i}c_{ij,ii'}c_{ij,ri}=\sigma^2_{\phi}\Lim{N\rightarrow\infty}N^{-1}\bmc^\textsc{t}_{i,ii'}\bmc_{i,ri}&i=r'\ ,\ i'\neq r\\ &\\
	\sigma^2_{\phi}\Lim{N\rightarrow\infty}N^{-1}\sum_{j'=1}^{n_{i'}}c_{i'j',ii'}c_{i'j',i'r'}=\sigma^2_{\phi}\Lim{N\rightarrow\infty}N^{-1}\bmc^\textsc{t}_{i',ii'}\bmc_{i',i'r'}& i\neq r'\ ,\  i'= r\\ &\\ 0 & \mbox{otherwise}
	\end{array}\right.\end{matrix}\nonumber
	\end{align}
	
	\noindent as given in \eref{covar}. To prove the theorem, we use the Cram\'{e}r-Wold device (Serfling \cite{serfling}, Theorem 1.5.2) to show that for every $\mathbf{d} = (d_{12}, \cdots, d_{1g}; \cdots; d_{g-1,g})^\textit{\textsc{t}}$ such that $\mathbf{d} \neq \mathbf{0}$ we have
	
	\[N^{-1/2}\mathbf{d}^\textsc{t}\bmS^* \xrightarrow{D} N(\mathbf{0}, \mathbf{d}^\textsc{t}\Sigma \mathbf{d}),\ \ as\ \  N\rightarrow \infty\]
	
\noindent Note that we can write
	
	\begin{align}
W=	\sum_{i=1}^{g-1}\sum_{i'=i+1}^{g} d_{ii'}S_{ii'}^*
	=&\sum_{i=1}^{g-1}\sum_{i'=i+1}^{g} d_{ii'}[-\sum_{j=1}^{n_i}c_{ij,ii'}t(Y_{ij})-\sum_{j'=1}^{n_{i'}}c_{i'j',ii'}t(Y_{i'j'})]\nonumber\\
	=&\sum_{j=1}^{N_1}(-\sum_{i'=2}^g d_{1i}c_{1j,1i'})t(Y_{1j})\nonumber\\
	&\ +\sum_{i=2}^{g-1}\sum_{j=1}^{n_i}(-\sum_{r=1}^{i-1} d_{ri}c_{ij,ri}-\sum_{i'=i+1}^g d_{ii'}c_{ij,ii'})t(Y_{ij})\nonumber\\
	&\ +\sum_{j=1}^{n_g}(-\sum_{r=1}^{g-1} d_{rg}c_{gj,rg})t(Y_{gj})\nonumber\\
	=&\sum_{i=1}^{g}\sum_{j=1}^{n_i}(-\sum_{r=1}^{i-1} d_{ri}c_{ij,ri}-\sum_{i'=i+1}^g d_{ii'}c_{ij,ii'})t(Y_{ij})\label{w}
	\end{align}
	where $d_{01}=0$, and $d_{g(g+1)}=0$.

\noindent 	Since $Y_{ij}$'s are independent, then following  H\'{a}jek and \v{S}id\'{a}k \cite{hajek} (Theorem v.1.2, p. 153), $N^{-1}W$ is asymptotically normally distributed if

\begin{equation} A_N/B_N \longrightarrow\infty,\ \ \mbox{if}\ \  N \longrightarrow\infty \label {AnBn}\end{equation}
where
\[A_N={\sum_{i=1}^{g}\sum_{j=1}^{n_i}(\sum_{r=1}^{i-1}d_{ri}c_{ij,ri}+\sum_{i'=i+1}^gd_{ii'}c_{ij,ii'})^2}\] \[B_N={\max_i\max_j(\sum_{r=1}^{i-1}d_{ri}c_{ij,ri}+\sum_{i'=i+1}^gd_{ii'}c_{ij,ii'})^2}\]

\noindent 	Using the Cauchy-Schwarz inequality and following assumption \eref{Noether}, we have 
	\begin{align} \lim_{N\rightarrow\infty}h_{ij,i'j'}&=\lim_{N\rightarrow\infty}\bmx^\textsc{t}_{ij}(X^\textsc{t}_{1,ii'}X_{1,ii'})^{-1}\bmx_{i'j'}\nonumber\\ &\leq \lim_{N\rightarrow\infty}\bmx^\textsc{t}_{ij}(X^\textsc{t}_{1,ii'}X_{1,ii'})^{-1}\bmx_{ij}\bmx^\textsc{t}_{i'j'}(X^\textsc{t}_{1,ii'}X_{1,ii'})^{-1}\bmx_{i'j'}=0 \label{hijipjp}\end{align}  for 	$i$, $i'=1,\ldots,g$, $j$, $j'=1,\ldots, n_i (n_{i'})$.
	
\noindent 	In order to \eref{AnBn}, we take a closer look at the expression of the general term from definition of $c_{ij,ii'}$ and $c_{i'j,ii'}$ in \eref{cijiip} and \eref{cipjpiip} and using \eref{hijipjp} we have

\begin{align*}
A_N &={\sum_{i=1}^{g}\sum_{j=1}^{n_i}(\sum_{r=1}^{i-1}d_{ri}c_{ij,ri}+\sum_{i'=i+1}^gd_{ii'}c_{ij,ii'})^2}\\
&=\sum_{i=1}^g\sum_{j=1}^{n_i}[\sum_{r=1}^{i-1}\frac{n_i}{n_r+n_i} d_{ri}\{\sum_{k=1}^{n_r}h_{rk,ij}+\frac{n_r}{n_i}
-\frac{n_r}{n_i}\sum_{k=1}^{n_i}h_{ik,ij}\}\\
&\hspace{.8in} {-\sum_{i'=i+1}^g\frac{n_i}{n_i+n_{i'}} d_{ii'}\{\frac{n_{i'}}{n_i}-\frac{n_{i'}}{n_i}\sum_{k=1}^{n_i}h_{ik,ij}+\sum_{k=1}^{n_{i'}}h_{i'k,ij}\}]^2}\\
&=\sum_{i=1}^g\sum_{j=1}^{n_i}[\sum_{r=1}^{i-1}\frac{\lambda_i}{\lambda_r+\lambda_i} d_{ri}\{o(n_r)+\frac{\lambda_r}{\lambda_i}(1-o(n_i))\}\\
&\hspace{.8in}  -\sum_{i'=i+1}^g\frac{\lambda_i}{\lambda_i+\lambda_{i'}} d_{ii'}\{\frac{\lambda_{i'}}{\lambda_i}(1-o(n_i))+o(n_{i'})\}]^2
\end{align*}
and
\begin{align*}
B_N &=\max_i\max_j(\sum_{r=1}^{i-1}d_{ri}c_{ij,ri}+\sum_{i'=i+1}^gd_{ii'}c_{ij,ii'})^2\\ &= \max_{i,j}[\sum_{r=1}^{i-1}\frac{n_i}{n_r+n_i} d_{ri}\{\sum_{k=1}^{n_r}h_{rk,ij}+\frac{n_r}{n_i}-\frac{n_r}{n_i}\sum_{k=1}^{n_i}h_{ik,ij}\}\\
&\hspace{.8in} {-\sum_{i'=i+1}^g\frac{n_i}{n_i+n_{i'}} d_{ii'}\{\frac{n_{i'}}{n_i}-\frac{n_{i'}}{n_i}\sum_{k=1}^{n_i}h_{ik,ij}+\sum_{k=1}^{n_{i'}}h_{i'k,ij}\}]^2}\\
& \max_{i,j}[\sum_{r=1}^{i-1}\frac{\lambda_i}{\lambda_r+\lambda_i} d _{ri}\{o(n_r)+\frac{\lambda_r}{\lambda_i}(1-o(n_i))\}\\
&\hspace{.8in} -\sum_{i'=i+1}^g\frac{\lambda_i}{\lambda_i+\lambda_{i'}} d_{ii'}\{\frac{\lambda_{i'}}{\lambda_i}(1-o(n_i))+o(n_{i'})\}]^2
\end{align*}

\noindent Since $\lim_{N\rightarrow\infty}({n_i}/{N}) =\lambda_i <\infty$ it follows that $\lim_{N\rightarrow\infty} (A_N/B_N) \longrightarrow \infty$.  This completes the proof.

%%%%%%%%%%%%%%%%%%%%%%%%%%%%%%%%%%%%%%%%%%%%%%%%%%%%%%%%%%%%%%%%%%%%%%%%%%%%%%%%%%%%%%%%%%%%%%%%%%%%%%%%%%%%%

\vspace{1cm} 
\noindent  \textit{Proof of Theorem 2}

\noindent Define \[S_{0,ii'}(\bmbeta)=\bmx^\textsc{t}_{0,ii'}a[\bmRipi(\bmbeta)],\ \ \mbox{and}\ \ \bmS^\textsc{t}_{1,ii'}(\bmbeta) = X^\textsc{t}_{1,ii'}a[\bmRipi(\bmbeta)], \  1\leq i<i'\leq g\]
 and let \[\bmS_{ii'}(\bmbeta)=X_{ii'}^\textsc{t}a[\bmRipi(\bmbeta)]=(S_{0,ii'}(\bmbeta),\bmS^\textsc{t}_{1,ii'}(\bmbeta))^\textsc{t}\]

\noindent It follows from Chiang and Puri \cite{chiang} (Lemas 3.1 \& 3.2) that
\begin{equation}
N^{-1/2}S_{0,ii'}(\hat{\bmbeta})=N^{-1/2}[1,-\bmx^\textsc{t}_{0,ii'}\X1ii(X^\textsc{t}_{1,ii'}\X1ii)^{-1}]\bmS_{ii'}(\bmbeta)+o_p(1), as\ N\rightarrow\infty \label{s0iip}
\end{equation}

\noindent From Mansouri \cite{mansouri15}(equation (A3), p. 673), we have that
\begin{equation}
N^{1/2}\hat{\theta}_{ii'}(\hat{\bmbeta})=N^{1/2}[\bmx^\textsc{t}_{0,ii'}(I-\H1ii)\bmx_{0,ii'}]^{-1}S_{0,ii'}(\hat{\bmbeta})+o_p(1),\ as\ N\rightarrow\infty \label{thetaiipapprox}
\end{equation}

\noindent where $\hat{\theta}_{ii'}(\hat{\bmbeta})$ is given by \eref{thetaiip}. Then following (\ref{s0iip}) and (\ref{thetaiipapprox}) we get

\begin{align} N^{1/2}\hat{\theta}_{ii'}(\hat{\bmbeta})&=N^{-1/2}\sigma_{ii'}^{-1}[1,-\bmx^\textsc{t}_{0,ii'}\X1ii(X^\textsc{t}_{1,ii}\X1ii)^{-1}]\bmS_{ii'}(\bmbeta)+o_p(1)\nonumber \\
&= N^{-1/2}\sigma^{-1}_{ii'}S^*_{ii'}(\bmbeta)+o_p(1)\ as\ N\rightarrow\infty \label{thetaiips*}\end{align}

\noindent where $\sigma_{ii'}$ and $S^*_{ii'}(\bmbeta)$ are given by (\ref{sigmaiip}) and \eref{s*iip}, respectively.

\noindent Consider $T[\bmR_{ii'}(\hat{\bmbeta})]$\ as given in \eref{Tiip}. Using \eref{thetaiip} and \eref{thetaiips*} we get

\begin{align}
T[\bmR_{ii'}(\hat{\bmbeta})]  &=  \{\hat{\sigma}^2_{\phi}{\bmx^\textsc{t}_{0,ii'}}(I-H_{ii'})\bmx_{0,ii'}\}^{-1/2}\bmx^\textsc{T}_{0,ii'}(I-H_{ii'})a[\bmRipi(\hat{\bmbeta})]\nonumber\\
&=\{\hat{\sigma}^2_{\phi}\}^{-1/2}\{\bmx^\textsc{t}_{0,ii'}(I-H_{ii'})\bmx_{0,ii'}\}^{1/2}\hat{\theta}_{ii'}(\hat{\bmbeta})\nonumber\\
&= \{\sigma^2_{\phi}\sigma_{ii'}\}^{-1/2}  N^{-1/2} S^*_{ii'}(\bmbeta)
 + o_p(1)\ \ \ \ \ \ \ \ as\ N\rightarrow\infty \label{Tiips*}\end{align}
 
\noindent Hence we have

\[\bmT[\bmR(\hat{\bmbeta})]=\sigma^{-1}_\phi \Delta (N^{-1/2} \bmS^*(\bmbeta)) + o_p(1)\ \ \mbox{as}\ \ N\rightarrow \infty \]

\noindent
where $\bmT[\bmR(\hat{\bmbeta})]$ is given by \eref{T} in which we had suppressed its dependence on $\hat{\bmbeta}$, $\bmS^*(\bmbeta)$ is defined in Lemma 1, and $\Delta$ is a diagonal matrix given by \[\Delta = \bigoplus_{1\leq i<i'\leq g}(\sigma^{-1/2}_{ii'})\] where $\bigoplus$ is the Kronecker sum.

\noindent Therefore the limiting distribution of $\bmT[\bmR(\hat{\bmbeta})]$ under $H_0$ is a multivariate $g(g-1)/2$ dimensional normal distribution with mean $\mathbf{0}$ and correlation matrix $C$ as defined in Theorem \ref{th:2} and can be written as

\[C=(\sigma^2_\phi)^{-1} \Delta \Sigma \Delta\] where $\Sigma$ is given by \eref{covar}.

%%%%%%%%%%%%%%%%%%%%%%%%%%%%%%%%%%%%%%%%%%%%%%%%%%%%%%%%%%%%%%%%%%%%%%%%%%%%%%%

\end{document}